\documentclass{article}
\usepackage{spconf}
\usepackage{amsmath}
\usepackage{wrapfig}
\usepackage{algorithmic}
\usepackage{array,subfig}
\usepackage{graphicx}
\usepackage{graphics}
\usepackage{lipsum}
\usepackage{caption}
\usepackage{multirow}
\usepackage{cite}
\usepackage[linesnumbered,ruled]{algorithm2e}
\usepackage{setspace}
\usepackage{array,booktabs}
\newcolumntype{Y}{>{\centering\arraybackslash}X}
\usepackage{float}
\usepackage{graphics}
\usepackage{graphicx}
\usepackage{caption}
\usepackage{threeparttable}
\usepackage{url}
\usepackage{xcolor}
\definecolor{uptxt}{rgb}{0,0,1}
\usepackage{setspace}
\usepackage{tikz}
\usepackage{url}
\usepackage{hyperref}
 
\usepackage{color, colortbl}
\usepackage{multirow}
\usepackage{multicol}
\usepackage{dblfloatfix}
\usepackage{pifont}
\usepackage{booktabs}
\usepackage{balance}
\definecolor{Gr}{gray}{0.8}

\ninept
\title{Boosting segmentation performance across datasets using histogram specification with application to pelvic bone segmentation}
\name{Prabhakara Subramanya Jois$^1$, Aniketh Manjunath$^2$, Thomas Fevens$^1$
\thanks{$\dagger$ This work was supported by Mitacs Accelerate Project-IT20604 and NSERC Grants RGPIN 04929 and RGPIN 06785, Canada.}
}
\address{$^1$Department of Computer Science and Software Engineering, Concordia University, Montr\'eal, Canada\\
$^2$Department of Computer Science, University of Southern California, Los Angeles, USA\\
Email: \{sp.subramanya, v.m.aniketh\}@gmail.com, thomas.fevens@concordia.ca}

\begin{document}
\ninept

\setlength{\belowcaptionskip}{-3pt}  
 \maketitle
\begin{abstract}  
Accurate segmentation of the pelvic CTs is crucial for the clinical diagnosis of pelvic bone diseases and for planning patient-specific hip surgeries. With the emergence and advancements of deep learning for digital healthcare, several methodologies have been proposed for such segmentation tasks. But in a low data scenario, the lack of abundant data needed to train a Deep Neural Network is a significant bottle-neck. In this work, we propose a methodology based on modulation of image tonal distributions and deep learning to boost the performance of networks trained on limited data. The strategy involves pre-processing of test data through histogram specification. This simple yet effective approach can be viewed as a style transfer methodology. The segmentation task uses a U-Net configuration with an EfficientNet-B0 backbone, optimized using an augmented BCE-IoU loss function. This configuration is validated on a total of 284 images taken from two publicly available CT datasets, TCIA (a cancer imaging archive) and the Visible Human Project. The average performance measures for the Dice coefficient and Intersection over Union are 95.7\% and 91.9\%, respectively, give strong evidence for the effectiveness of the approach, which is highly competitive with state-of-the-art methodologies. 

\end{abstract}


\section{Introduction}
\label{sec:intro} 
In recent years, due to the increase in the incidence of pelvic injuries from traffic-related accidents \cite{incidence}, pelvic bone diseases within the aging population, and sufficient access to computed tomography (CT) imaging, automated pelvic bone segmentation in CT has gained considerable prominence. The segmentation results assist physicians in the early detection of pelvic injury and help expedite surgical planning and reduce the complications caused by pelvic fractures \cite{intro}. In CT data, structures like the bone marrow and bone surface appear as dark and bright regions due to their low and high densities compared to the surrounding tissues. However, given the variations in image quality between different CT datasets, distinguishing bone structures from the image background becomes cumbersome and leads to erroneous segmentation outputs. These issues indicate a need for a novel solution to develop a simple yet effective methodology for the accurate segmentation of pelvic bones from varying CT data.

\begin{figure}[t]
\centering
$\begin{array}{cc}
\includegraphics[width=35mm,height = 35mm]{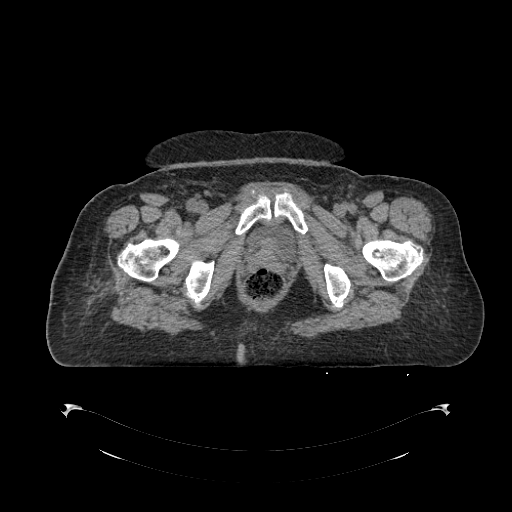}& \includegraphics[width=35mm,height = 35mm]{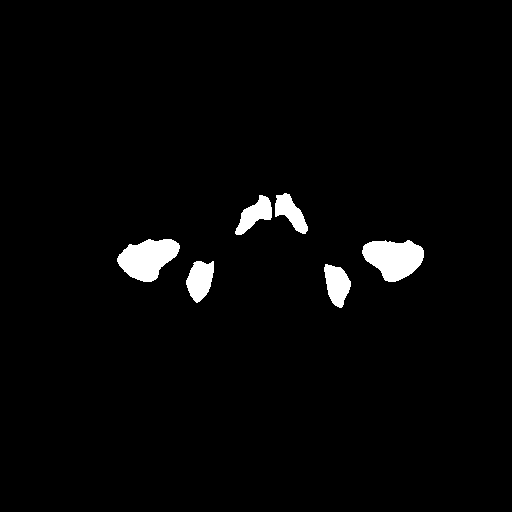}\\
\mbox{TCIA}&\mbox{(a1)}\\
\includegraphics[width=35mm,height = 35mm]{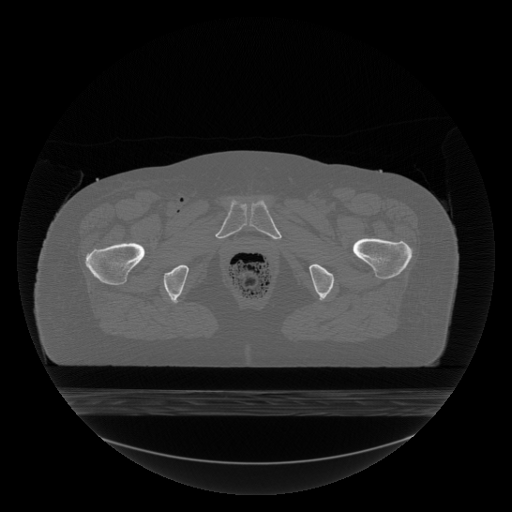}& \includegraphics[width=35mm,height = 35mm]{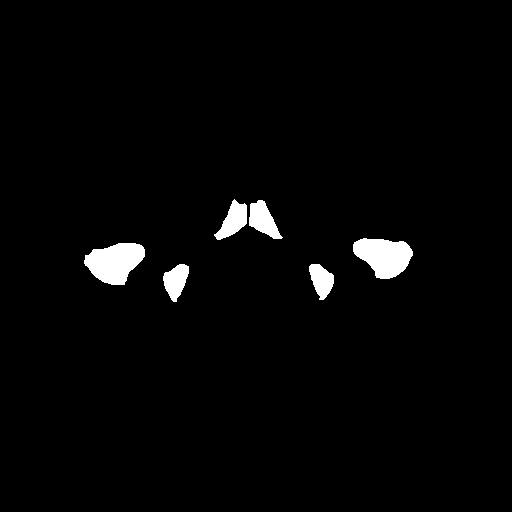}\\
\mbox{VHBD}&\mbox{(b1)}\\
\end{array}$
\caption{(a1) and (b1) -- illustrate the segmentation outputs, for input images from TCIA \cite{TCIA} and VHBD \cite{VHBD}, respectively.}
\label{fig1}
\end{figure}

\noindent 
\textbf{Contribution of this paper:} The key novelties of this work are as follows:
\begin{enumerate}
    \item introduction of an encoder-decoder network, trained on limited data, for high accuracy segmentation of pelvic bones
    \item boosting model performance on unseen data by employing histogram specification
\end{enumerate}
\noindent The exact details of the approach are deferred until Sec.~\ref{hist}. Fig.~\ref{fig1} illustrates the results of the proposed method.

\section{Prior Art}
\label{sec:prior}

\begin{figure*}
\centering
\includegraphics[width= 0.9\textwidth, height = 0.27\textwidth]{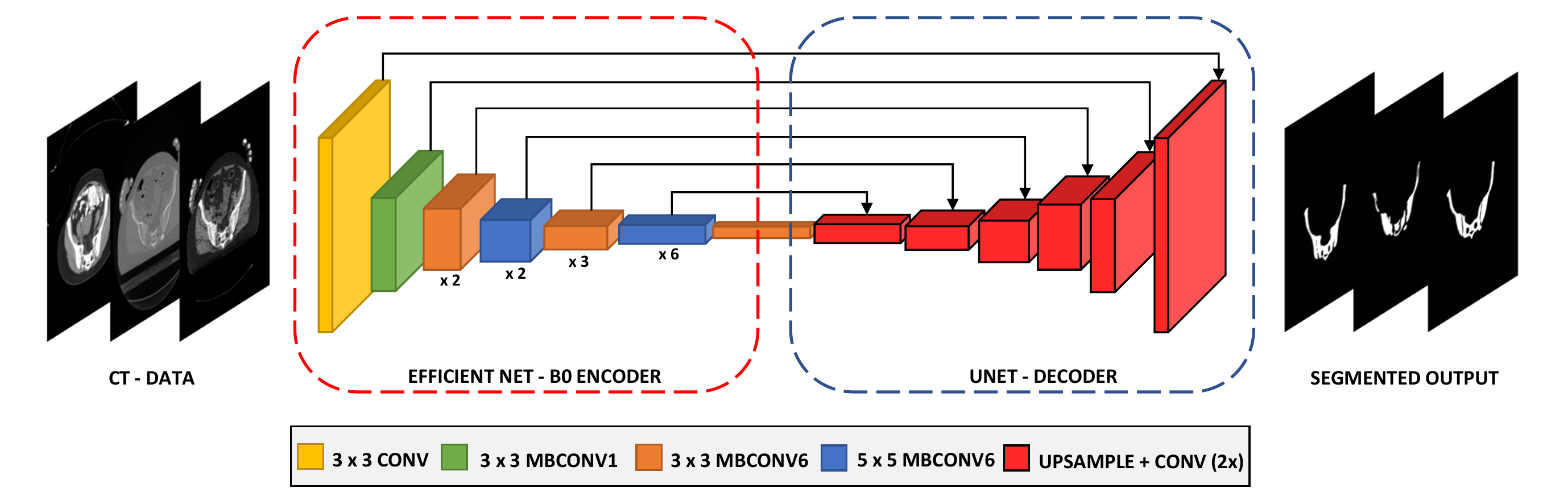}
\caption{Workflow of U-Net architecture with pre-trained backbone, detailing pelvic bone segmentation.}
\label{fig2}
\end{figure*}

Recent literature has seen many applications for the segmentation of the pelvis from CT imaging data. 
Traditional methods such as thresholding and region growth \cite{truc2008density}, deformable surface model \cite{kainmueller2008coupling}, and others, have been commonly used to perform bone segmentation. However, these approaches often suffer from low accuracy due to varying image properties such as intensity, contrast, and the inherent variations between the texture of the bone structures (bone marrow and surface boundary) and the surrounding tissues. To overcome these challenges, supervised methods such as statistical shape models (SSM) and atlas-based deep learning (DL) methods have made significant contributions to segmentation tasks. Wang {\em et al.~}\cite{wang2012multi, wang2017multi} suggested using a multi-atlas segmentation with joint label fusion for detecting regions on interest from CT images. Yokota {\em et al.~}\cite{yokota2013} showcased a combination of hierarchical and conditional SSMs for the automated segmentation of diseased hips from CT data.  Chu {\em et al.~}\cite{chu2015mascg} presented a multi-atlas based method for accurately segmenting femur and pelvis. Zeng {\em et al.~}\cite{zeng20173d} proposed a supervised 3D U-Net with multi-level supervision for segmenting femur in 3D MRI. Chen {\em et al.~}\cite{chen2017three} showcased a 3D feature enhanced network for quickly segmenting femurs from CT data. Chang {\em et al.~}\cite{chang2018accurate} proposed patch-based refinement on top of a conditional random field model for fine segmentation of healthy and diseased hips. Liu {\em et al.~}\cite{liu2020deep} used 3D U-Nets in two-stages (trained on approximately 270K images) with a signed distance function for producing bone fragments from image-stacks. In the following section, we discuss a new technique addressing accurate segmentation of the pelvis from CT images of varying qualities.

\section{Proposed Methodology}
\label{sec:methods}
The efficacy of using Encoder-Decoder architectures for designing high accuracy segmentation models for biomedical applications has been showcased in recent literature \cite{Unet,zeng20173d, liu2020deep}. We employ a similar architecture, with various encoder modules for feature extraction and a decoder module for semantic segmentation. The details of the encoder and decoder modules are explained in the following. 

\subsection{\textbf{Encoder Module}}
In simple terms, an encoder takes the input image and generates a high-dimensional feature vector aggregated over multiple levels. We deploy a choice of the following well-known architectures as the encoder module:

\subsubsection{ResNet}
Residual networks (ResNet) introduced residual mappings to solve the vanishing gradient problem in deep neural networks \cite{resnet34}. ResNets are easy to optimize and gain accuracy even with deeper models. 

\subsubsection{Inception V3}
Inception Networks are computationally efficient architectures, both in terms of the model parameters and their memory usage. Adapting the Inception network for different applications while ensuring that changes do not impede its computational efficiency is difficult. Inception V3 introduced various strategies for optimizing network with ease of model adaptation capabilities \cite{inceptionv3}.

\subsubsection{EfficientNet}
Conventional methods make use of scaling to increase the accuracy of the models. The models are scaled by increasing the depth/width of the network or using higher resolution input images. EfficientNet results from a novel scaling method that uses a compound coefficient to uniformly scale the network across all dimensions \cite{efficientb0}.


\subsection{\textbf{Decoder Module}}
The decoder module is responsible for generating a semantic segmentation mask using the aggregated high-dimensional features extracted by the encoder module. We make use of the popular U-Net model specially designed for medical imaging as the decoding module \cite{Unet}.

\subsection{Histogram Specification}
\label{hist}

Histogram specification, or histogram matching, is a traditional image processing technique \cite{cvbook} that matches the input image's histogram to a reference histogram. Histogram specification involves computing the cumulative distribution function (CDF) of histograms from both the target and the reference, following which a transformation function is obtained by mapping each gray level $[0, 255]$ from the target's CDF (input) to the gray level in the reference CDF. In this work, we construct the reference histogram by averaging over histograms from every image in the training set.   Using this technique as a pre-processing step for the test data serves an important purpose, as the distribution of the test data is converted to a similar form seen by the network during training. 

\section{Experimental Validation}
\label{sec:exper}
\begin{table}[!t]
\centering
\footnotesize
\caption{An overview of the datasets used in this work.}
\begin{tabular}{lllll}
\toprule
\multirow{2}{*}{\textbf{Dataset}} &\textbf{Resolution} & \textbf{Train-set} & \textbf{Val-set} & \textbf{Test-set} \\ 
 &\textbf{\# Images} & \textbf{(\%)} & \textbf{(\%)} & \textbf{(\%)} \\ 
\midrule
\multirow{2}{*}{\textbf{TCIA \cite{TCIA}}}&  512 x 512 & 407 & 58& 117 \\ 
&  582 & (70\%) & (10\%) & (20\%) \\ 
\midrule
\multirow{2}{*}{\textbf{VHBD \cite{VHBD}}}&  512 x 512 & \multirow{2}{*}{--} & \multirow{2}{*}{--} & 167 \\ 
&  167 &  & & (100\%) \\ 
\midrule
\multirow{2}{*}{\textbf{VHBD-2 \cite{VHBD}}}&  512 x 512 & 116 & 17 & 34 \\ 
&  167 & (70\%) & (10\%) & (20\%) \\ 
\bottomrule
\end{tabular}
\label{dataset}
\end{table}

\subsection{Datasets}
\label{sec:dataprep}
The input data preparation and label annotation were done using the tools from \textit{Image-J} software. A summary of TCIA--cancer imaging archive \cite{TCIA} and VHBD--Visible human project \cite{VHBD} datasets, image resolution, the number of images used in this study, and the respective data-splits for training-validation-testing, are shown in Table~\ref{dataset}.

\subsection{Performance Measures}
\label{sec:permer}
To quantify the quality of segmentation, we compute standard performance measures for segmentation tasks commonly used in literature, specifically, the mean Dice coefficient (mDice) and mean Intersection over Union (mIoU) \cite{dice,jaccard}. For a given segmentation output ($A$) and the ground truth ($B$), the Dice coefficient is given by 
$\text{Dice} =\frac{2\,|A\;\cap\;B|}{|A|\;+\;|B|}$, which can be interpreted as a weighted average of the precision and recall, and 
$\text{IoU} = \frac{|A\;\cap\;B|}{|A\;\cup\;B|}\text,$ (also known as Jaccard index) is commonly used for comparing the similarity between sets ($A$) and ($B$), while penalizing their diversity.

\subsection{Network Training}
The implementations used were based on the documentation from \cite{Y_pavel}. The models used \cite{resnet34, inceptionv3, efficientb0} were pre-trained on the Imagenet \cite{imagenet} dataset to improve the generalization capability on unseen data and achieve faster convergence. For the base-model, we use ResNet-34 \cite{resnet34} as the encoder and a U-Net decoder. We initialize the base-model with random weights (\textbf{rnwt}) and train without any data-augmentation (\textbf{noaug}) on images from \cite{TCIA}, using an Nvidia RTX 2070 GPU, and an ADAM optimizer with a learning rate of 0.001, momentum of 0.9 and a weight decay of 0.0001, for 40 epochs. We chose a 70\% : 10\% : 20\% split of the data (shown in the first row of Table~\ref{dataset}), where the 70\% was utilized for training and the 10\% of the data was utilized for validation. The remaining 20\% for testing was completely unseen during training. About 50 passes of random image batches, of size eight, from the training set, were used in each epoch. The model was then validated on the 10\% data to evaluate the performance based on binary-cross-entropy loss (\textbf{bce}) and record the corresponding weights. After training, the weights that gave the best performance on the validation set were selected for the base-model, which was then evaluated on the unseen test-sets, i.e., 20\% of \cite{TCIA} and 100\% of \cite{VHBD}, respectively, whose performance is showcased in the first row of Table~\ref{test}.

Extending beyond the base-model, data augmentation (\textbf{aug}) was performed using horizontal and vertical flips, affine transforms, image intensity modulation and blurring, for increasing training data size and to help reduce over-fitting. In addition, we try to find the best overall segmentation performance and generalization capability to completely unseen data, through further extension of the base-model with different configurations, using the following: 
\begin{itemize}
    \item \textit{encoder modules} using ResNet-34 \cite{resnet34}, Inception V3 \cite{inceptionv3} and EfficientNet-B0 \cite{efficientb0}, initialized with Imagenet weights (\textbf{imwt}) for transfer learning
    
    \item  \textit{re-configuration} of input data, or not, to the pre-trained model's format and its pre-processing functions (\textbf{ppr}), for extraction of better features 
    
    \item \textit{loss functions} like Dice loss (\textbf{dice)}, IoU loss (\textbf{iou}) and combined \textbf{bce-iou} loss, in place of \textbf{bce} loss, for propagating strong gradients for better optimization and learning

\end{itemize}

\begin{figure}[!t]
\centering
$\begin{array}{ccc}
\includegraphics[width=25mm,height = 25mm]{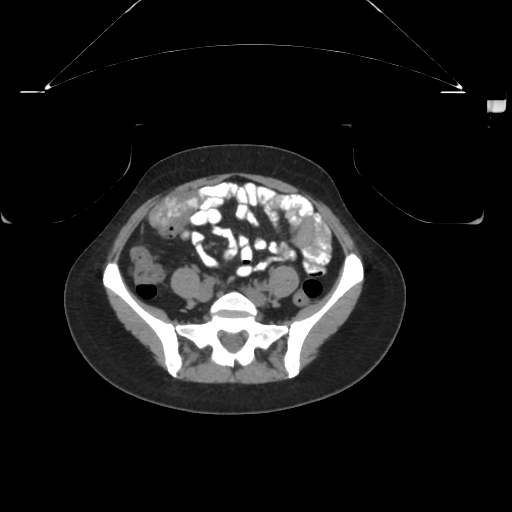}& \includegraphics[width=25mm,height = 25mm]{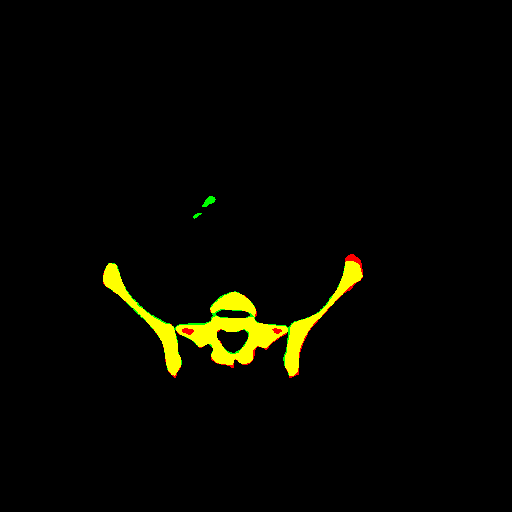}
& \includegraphics[width=25mm,height = 25mm]{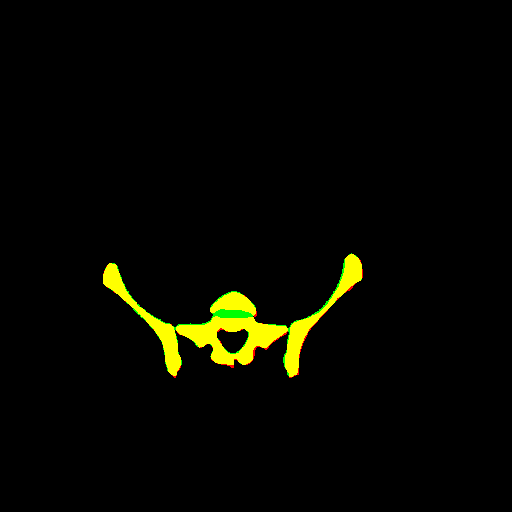}\\
\mbox{Input}&\mbox{(a)}&\mbox{(b)}\\
\end{array}$
\caption{Pelvic bone segmentation on TCIA data using: (a) Base U-Net with random weight initialization for ResNet-34 encoder, with no data-augmentation, optimized using BCE loss (\textbf{least performing}); and (b) fine-tuned U-Net with Imagenet weight initialization for EfficientNet-B0 encoder, with data-augmentation and input reconfiguration, optimized using combined BCE-IoU loss (\textbf{best performing}), are overlaid  onto the  binary ground-truth; yellow - TP; black - TN; green - FP; red - FN.}
\label{fig3}
\end{figure}

\begin{figure}[!t]
\centering
$\begin{array}{ccc}
\includegraphics[width=25mm,height = 25mm]{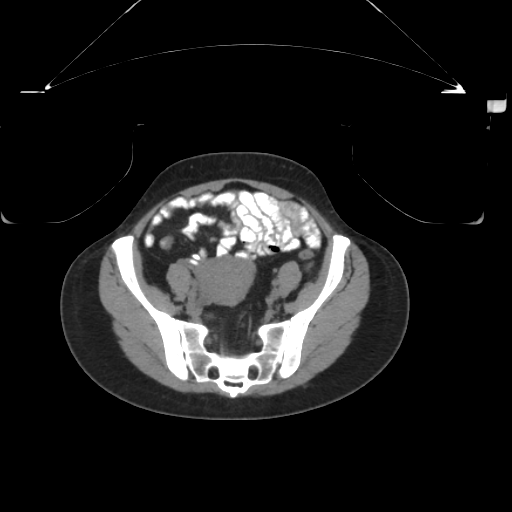}& \includegraphics[width=25mm,height = 20mm]{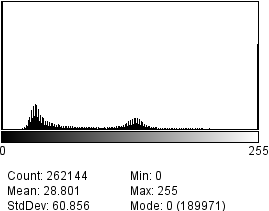}
& \includegraphics[width=25mm,height = 25mm]{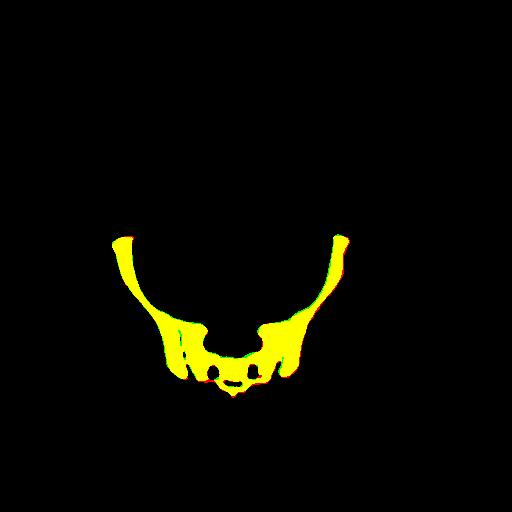}\\
\mbox{TCIA}&\mbox{(a1)}&\mbox{(a2)}\\
\includegraphics[width=25mm,height = 25mm]{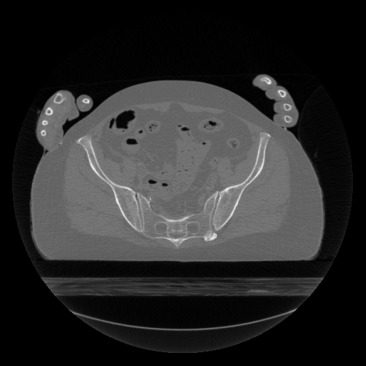}& \includegraphics[width=25mm,height = 20mm]{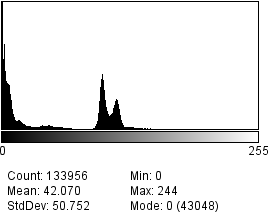}
& \includegraphics[width=25mm,height = 25mm]{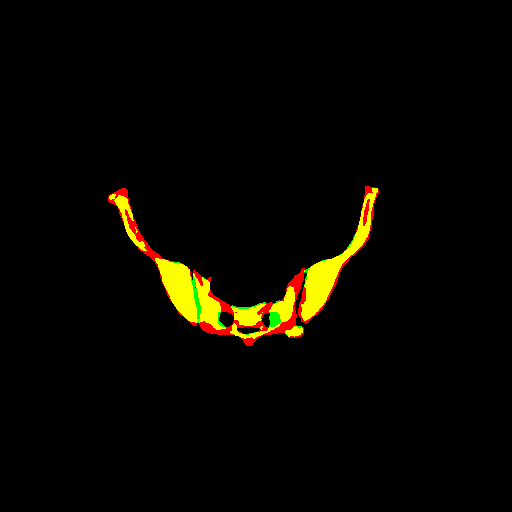}\\
\mbox{VHBD (target)}&\mbox{(b1)}&\mbox{(b2)}\\
\includegraphics[width=25mm,height = 25mm]{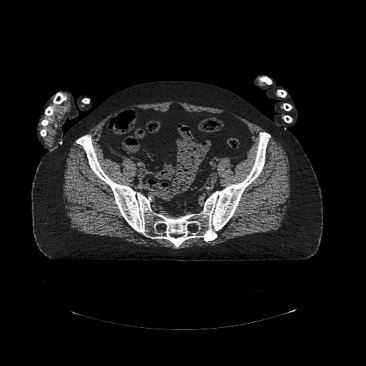}& \includegraphics[width=25mm,height = 20mm]{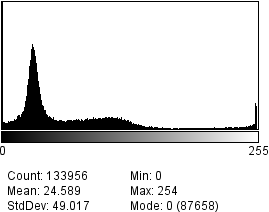}
& \includegraphics[width=25mm,height = 25mm]{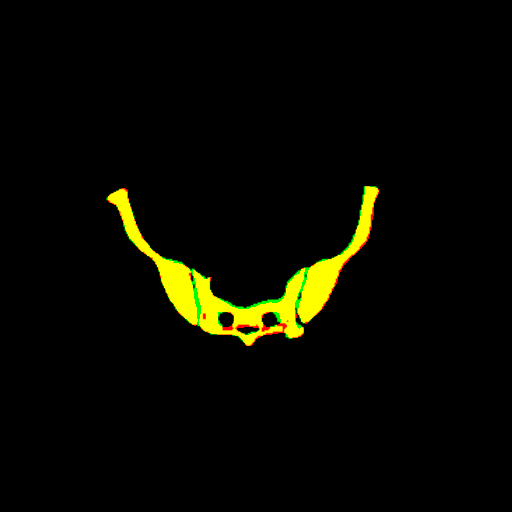}\\
\mbox{H-VHBD}&\mbox{(c1)}&\mbox{(c2)}\\
\end{array}$
\caption{Performance in segmentation with histogram specification: (a1-c1) show the respective histograms of the input images; (a2-c2) show the pelvic bone segmentations overlaid on the ground-truth; and (b2-c2) decisively show the improvement in segmentation from matching target's histogram to the reference. yellow - TP; black - TN; green - FP; red - FN.}
\label{fig4}
\end{figure}

\begin{table*}[!ht]
\centering
\footnotesize
\caption{Performance comparison of different U-Net configurations for pelvic bone segmentation on unseen data from \textbf{TCIA}, \textbf{VHBD}, and \textbf{H-VHBD}, i.e., VHBD after histogram specification.}
\begin{threeparttable}
\vspace{-3mm}
\begin{center}
\footnotesize
\begin{tabular}{p{3.5cm}p{1.7cm}p{1.7cm}p{1.7cm}p{1.7cm}p{1.7cm}p{1.7cm}}
\toprule
\multirow{2}{*}{\textbf{U-Net Configurations}}& \multicolumn{2}{c}{\textbf{TCIA}} & \multicolumn{2}{c}{\textbf{VHBD}} & \multicolumn{2}{c}{\textbf{H-VHBD}}  \\ \cmidrule{2-3}\cmidrule{4-5} \cmidrule{6-7} 
 &\textbf{mIoU} &\textbf{mDice} & \textbf{mIoU} & \textbf{mDice} & \textbf{mIoU} & \textbf{mDice}\\ 
\toprule
\textbf{Res34-rnwt-noaug-bce}& 
0.788 $\pm$ 0.033 &
0.867 $\pm$ 0.025 & 
0.131 $\pm$ 0.033 & 
0.186 $\pm$ 0.037 &
\cellcolor{Gr} 0.774 $\pm$ 0.021 & 
\cellcolor{Gr}0.865 $\pm$ 0.015  \\ 
\midrule
\textbf{Res34-imwt-aug-bce}& 
0.919 $\pm$ 0.007 & 
0.957 $\pm$ 0.004 & 
0.746 $\pm$ 0.028 & 
0.840 $\pm$ 0.021 & 
\cellcolor{Gr} 0.900 $\pm$ 0.004 &
\cellcolor{Gr} 0.947 $\pm$ 0.002\\ 
\midrule
\textbf{Res34-imwt-aug-dice}&  
0.925 $\pm$ 0.007 & 
0.960 $\pm$ 0.004 &  
0.523 $\pm$ 0.039 & 
0.645 $\pm$ 0.038 & 
\cellcolor{Gr} 0.907 $\pm$ 0.005 & 
\cellcolor{Gr} 0.951 $\pm$ 0.002\\
\midrule
\textbf{Res34-imwt-aug-bce-iou}& 
0.922 $\pm$ 0.007 & 
0.959 $\pm$ 0.004 & 
0.790 $\pm$ 0.018 & 
0.877 $\pm$ 0.012  & 
\cellcolor{Gr} 0.906 $\pm$ 0.004 & 
\cellcolor{Gr}  0.950 $\pm$ 0.002\\ 
\midrule
\textbf{IncepV3-imwt-aug-bce}& 
0.876$\pm$ 0.012 & 
0.932 $\pm$ 0.007 & 
0.663 $\pm$ 0.026 & 
0.784 $\pm$ 0.019 & 
\cellcolor{Gr} 0.906 $\pm$ 0.006 & 
\cellcolor{Gr} 0.950 $\pm$ 0.003 \\ 
\midrule
\textbf{IncepV3-ppr-imwt-aug-bce}& 
0.921 $\pm$ 0.011& 
0.957 $\pm$ 0.007 & 
0.808 $\pm$ 0.014 & 
0.890 $\pm$ 0.009 &  
\cellcolor{Gr} 0.913 $\pm$ 0.005 & 
\cellcolor{Gr} 0.954 $\pm$ 0.002\\ 
\midrule
\textbf{EffiB0-imwt-aug-bce}& 
0.923 $\pm$ 0.007 & 
0.959 $\pm$ 0.004  & 
0.835 $\pm$ 0.015 & 
0.906 $\pm$ 0.010 & 
\cellcolor{Gr} 0.901 $\pm$ 0.006 & 
\cellcolor{Gr} 0.947 $\pm$ 0.003\\ 
\midrule
\textbf{EffiB0-ppr-imwt-aug-bce-iou}& 
\textbf{0.924 $\pm$ 0.008} & 
\textbf{0.960 $\pm$ 0.004}  & 
\textbf{0.836 $\pm$ 0.011} & 
\textbf{0.909 $\pm$ 0.006}  & 
\cellcolor{Gr} \textbf{0.914 $\pm$ 0.005} & 
\cellcolor{Gr} \textbf{0.955 $\pm$ 0.002}\\ 
\midrule
\textbf{ABLATION STUDY (\ding{168})}& 
0.913 $\pm$ 0.007 & 
0.954  $\pm$ 0.004  & 
0.679 $\pm$ 0.047 & 
0.801 $\pm$ 0.032 & 
\cellcolor{Gr} 0.873 $\pm$ 0.013 & 
\cellcolor{Gr} 0.931 $\pm$ 0.007\\ 
\toprule
\end{tabular}
\begin{tablenotes}
\item[*]\textbf{Encoder module} - Res34, IncepV3, EffiB0 are ResNet-34, Inception Net-V3, EfficientNet-B0, respectively.
\item[*]\textbf{Encoder Weights} - rnwt and imwt are random weights and Imagenet weights, respectively.
\item[*]\textbf{Augmentation} - aug and noaug means training with and without data-augmentation, respectively.
\item[*]\textbf{Loss} - bce, dice, iou are the Binary Cross Entropy loss, Dice Loss and IoU loss, respectively .
\item[*]\textbf{ppr}- configure input to the pre-trained backbone's format.\\ \vspace{-2.5mm}
\item[*]\colorbox{Gr}{Grey background}- indicates improvement due to histogram-specification based pre-processing.
\end{tablenotes}
\end{center}
\end{threeparttable}
\label{test}
\end{table*}

\subsection{Results}
The detailed comparisons of the different U-Net configurations' segmentation performance on test-sets with 95\% confidence intervals are shown in Table~\ref{test}. The segmentation outputs from the \textit{least-performing} (base-model) and \textit{best-performing} (fine-tuned U-Net with Imagenet weight initialization for EfficientNet-B0 encoder \cite{efficientb0}, with data-augmentation and input re-configuration, optimized using combined BCE-IoU loss) DL models are showcased in Fig.~\ref{fig3} (a) \&(b). The predicted outputs are overlaid onto the ground-truth and color-coded (yellow - TP; black - TN; green - FP; red - FN) for visualizing the quality of segmentation. The results shown in Fig.~\ref{fig4}(b2) \&(c2) illustrate the desired effect on segmentation due to histogram specification. The reduction in the number of pixels labeled as FPs \& FNs, and improvement in number of TPs from the overlays decisively show the significance of pre-processing test-data, which clearly boosts the model's segmentation performance. Furthermore, the comparitive results tabulated in the last two columns of Table~\ref{test} give strong evidence for the success of the proposed methodology on all the specified model configurations.

On analyzing the data shown in Table~\ref{finres}, the proposed methodology's overall performance on the test-sets surpassed several state-of-the-art techniques that were trained on similarly sized datasets, with the exception of Liu {\em et al.~}\cite{liu2020deep} who performed training on approximately 270,000 images. Since data drives any model, the proposed methodology (trained only on 407 images) shows room for further improvement in segmentation under the availability of larger datasets.

\begin{table} [!t]
\centering
\footnotesize
\caption{Overall performance comparison for pelvic bone segmentation with state-of-the-art techniques.}
\begin{threeparttable}
\begin{tabular}{llll}
\toprule
{\textbf{Methodology}} &\textbf{Dataset}  & \textbf{mIoU} & \textbf{mDice}  \\ 
 &(\# Images) &  &   \\ 
\midrule
{Yokota et al. \cite{yokota2013}}& Private  (100) & --- & 0.928    \\ 
\midrule
{Chu et al. \cite{chu2015mascg}}& Private (318)& --- &  0.941  \\ 
\midrule
{Chang et al. \cite{chang2018accurate}}& Private ($\sim$3420)  & --- &  0.949  \\ 
\midrule
{Liu et al. \cite{liu2020deep}}& DS$\ddagger$($\sim$63K) & --- & \textbf{0.984}  \\ 
\midrule
{\textbf{Proposed method}}& TCIA,VHBD (284)  & 0.919 &  0.957 \\ 
\bottomrule 
\end{tabular}
\begin{tablenotes}
\item[]\scriptsize{(DS$\ddagger$) KITS19, CERVIX, ABDOMEN, MSD T10, COLONOG, CLINIC; {Train:Test $\approx$ 270K: 63K;  K = $10^{3}$}}
\end{tablenotes}
\end{threeparttable}
\label{finres}
\end{table}

\subsection{Ablation Study}
Images from \cite{TCIA,VHBD}, with the data splits shown in rows 1 and 3 of Table~\ref{dataset}, are used for training. The \textit{best-model} was trained on the joint data whose test-data performance is shown in Table~\ref{test} (\ding{168}). The results showed that training the model on joint data degrades the performance on both datasets. The data imbalance and the varying image tonal distributions play a significant role in influencing the segmentation performance. And by using the proposed methodology, the model overcomes data imbalance and generalizes well to unseen datasets, which boosts its overall segmentation performance.

\section{Conclusion}
To sum up, in this work,  we presented a novel methodology for the automated segmentation of pelvic bones from axial CT images. We addressed the unmet need for superior pelvic bone segmentation methodology for images with varying properties by using histogram specification. This simple yet powerful approach of pre-processing the test-data improved segmentation performance by a significant margin, with the quantitative results confirming its validity.  Through our approach, the encoder-decoder configuration overcame a significant hurdle of varying intensity distributions in CT images, which led to superior segmentation quality. Moreover, after validating the results on publicly available TCIA and VHBD datasets, the proposed methodology has been shown to be highly competent with-respect-to existing state-of-the-art techniques.

Through this study, we saw that, although deep learning has pushed the limits for image processing applications, traditional image processing techniques are not necessarily obsolete and that combining the two approaches can lead to superior performance in segmentation.   






\bibliographystyle{IEEEbib}
\balance
\bibliography{My_ref}

\end{document}